\documentclass[superscriptaddress,prl,twocolumn,amsmath,twocolumn,amssymb]{revtex4-1}

\usepackage{amssymb,xcolor}
\usepackage{epsfig,amsmath}

\newcommand{\ket}[1]{|#1\rangle}

\newcommand{\citeFig}[1]{Fig.~\ref{#1}}

\newcommand{\cha}[1] {\textcolor{black}{#1}}

\newcommand{\ubristol}{Centre for Quantum Photonics, H. H. Wills Physics Laboratory and Department of Electrical and Electronic Engineering, University of Bristol, Merchant Venturers Building, Woodland Road, Bristol BS8 1UB, UK.}
\newcommand{\ifn}{Istituto di Fotonica e Nanotecnologie, Consiglio Nazionale delle Ricerche,
Piazza L. da Vinci, 32, I-20133 Milano, Italy}
\newcommand{\polimi}{Dipartimento di Fisica, Politecnico di Milano, Piazza L. da Vinci, 32, I-20133 Milano, Italy}
\newcommand{\bristolphysio}{Microvascular Research Labs, School of Physiology and Pharmacology, Preclinical Vet Building, University of Bristol BS2 8EJ UK}

\begin{document}

%\title{Quantum biosensing}
\title{Measuring protein concentration with entangled photons}

\author{Andrea Crespi}
\affiliation{\ifn}
\affiliation{\polimi}
\author{Mirko Lobino}
\affiliation{\ubristol}
\author{Jonathan C. F. Matthews}
\affiliation{\ubristol}
\author{Alberto Politi}
\affiliation{\ubristol}
\author{Chris~R.~Neal}
\affiliation{\bristolphysio}
\author{Roberta Ramponi}
\affiliation{\ifn}
\affiliation{\polimi}
\author{Roberto Osellame}
\affiliation{\ifn}
\affiliation{\polimi}
\author{Jeremy L. O'Brien}
\email{Jeremy.OBrien@bristol.ac.uk}
\affiliation{\ubristol}

\begin{abstract}
Optical interferometry is amongst the most sensitive techniques for precision measurement. By increasing the light intensity a more precise measurement can usually be made. However, in some applications the sample is light sensitive.  By using entangled states of light the same precision can be achieved with less exposure of the sample. This concept has been demonstrated in measurements of fixed, known optical components. Here we use two-photon entangled states to measure the concentration of the blood protein bovine serum albumin (BSA) in an aqueous buffer solution. We use an opto-fluidic device that couples a waveguide interferometer with a microfluidic channel. These results point the way to practical applications of quantum metrology to light sensitive samples.
\end{abstract}

\maketitle

Even the most advanced sensors are bound by a hard limit in precision---the shot noise or standard quantum limit (SQL) that arises from statistical fluctuations.
In a conventional optical interferometer for example the precision with which an unknown optical phase $\phi$ can be measured is limited to $\delta\phi=1/\sqrt{\textrm{N}}$, where N is the (average) number of photons used to probe $\phi$ \cite{gi-sci-306-1330,wa-nat-429-158,mitchell2004pol,nagata2007bsq,hi-nat-450-393,ma-nphot-3-346}. Increasing $N$ is usually possible, by increasing laser power for example. However, in some scenarios the practical limits of laser power are reached and increasing the integration time will reduce the bandwidth of the measurement below that required---gravity wave interferometers are a key example \cite{go-nphys-4-472}.
In other scenarios the sample to be measured may be sensitive to light, such that one would like to minimise the photon flux or total number of photons that the sample is exposed to in order to reach the required precision; put another way, one wishes to gain the maximum information allowable by the laws of physics for the given perturbation of the sample.
It is this latter scenario that we are focussed on here.
By harnessing quantum superposition and entanglement the SQL can be overcome---quantum metrology enables the more fundamental Heisenberg limit of precision, $\delta\phi=1/\mathrm{N}$, to be reached \cite{gi-sci-306-1330}. However, practical applications of quantum metrology require that samples of interest are integrated with quantum optical circuits. 

We use the optofluidic device shown in Fig. 1, consisting of a microfluidic channel that passes through one arm of a Mach-Zehnder interferometer (MZI), fabricated by femtosecond laser microfabrication \cite{mo-nphoton-1-106,gattass2008flm,osellame2009ioe}.
This device combines the stability of integrated optics for high visibility quantum and classical interference \cite{po-sci-320-646,la-apl-97-211109} %,ma-nphot-3-346,sm-oe-17-13516,po-sci-325-1221,pe-sci-329-1500,sa-prl-105-200503,,na-apl-96-211101}
with high precision handling of fluid samples %chemical and biological samples
\cite{whitesides2006ofm,mo-nphoton-1-106}.
When a solution is fed into the microfluidic channel, any relative phase shift of light (and thereby concentration-dependent refractive index) in the sensing arm with respect to that acquired in the reference arm can be estimated from the interference fringes.

The period of the interference fringes and the measurement precision $\delta\phi$ depend on the particular state of light used to probe the sample. For classical states of light $\delta\phi\ge1/\sqrt{\textrm{N}}$, the SQL. Going beyond this limit requires quantum states of light; the canonical example is the NOON state $(|\mathrm{N}0\rangle+|0\mathrm{N}\rangle)/\sqrt{2}$, which can achieve super-sensitivity and saturate the Heisenberg limit $\delta\phi=1/\mathrm{N}$. Here, super-resolution results in a fringe periodicity that is $1/\mathrm{N}$ times shorter than the one obtained with classical  light \cite{gi-sci-306-1330}.
The generation and detection of NOON states with large $N$ is an active area of research \cite{af-sci-sci-328-879,ma-arxiv-1005.5119,ca-prl-99-163604}.
We test the operation of our device for the $N=2$ NOON state, which enables super-sensitity and super-resolution, and can be generated from two single photons input into a beamsplitter \cite{ra-prl-65-1348}. 
Photon losses in the interferometer reduce the measurement sensitivity, making NOON states non-optimal in general \cite{ru-pra-75-053805}.
However, for two photons, the NOON state is optimal \cite{ho-prl-71-1355,do-prl-102-040403}, and we calculate that for the MZI configuration the best obtainable performance is achieved when input and output beam splitters have 50:50 splitting ratios (see Supplementary Information).

\begin{figure*}
		\centering
		\includegraphics[width=16cm]{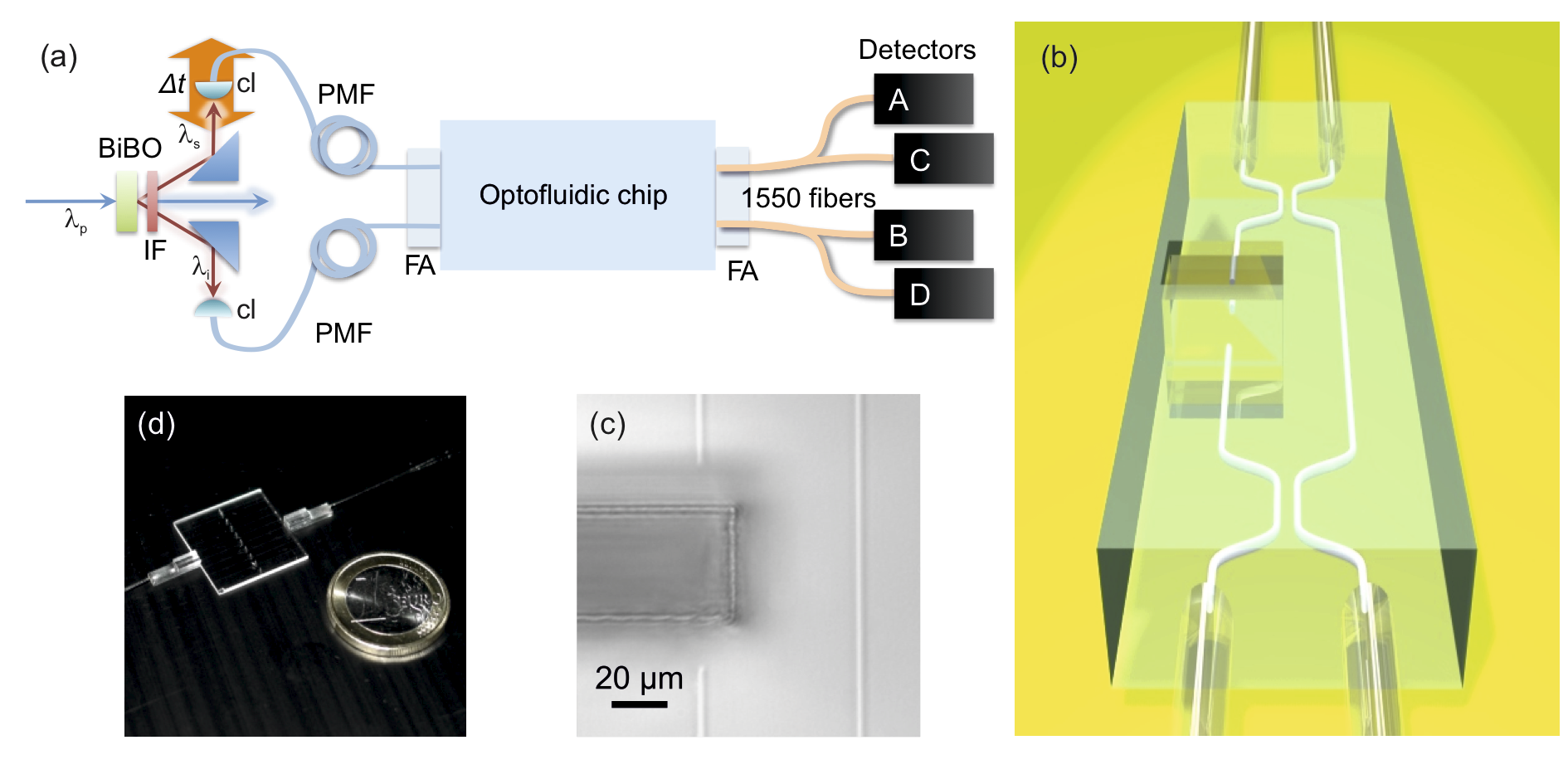}
		\caption{Quantum metrology in an optofluidic device. (a) Schematic of the experimental setup: A pump laser at $\lambda_p=392.5$~nm generates pairs of downconverted photons at $\lambda_s$~=~$\lambda_i$~=~785~nm in a Bismuth Borate (BiBO) crystal. IF: interference filter, cl: collection lenses, PMF: polarization maintaining fibers, FA: fiber array. (b) Schematic of the Mach-Zehnder interferometer (MZI) interfaced to the microchannel. The fluidic channel has rectangular cross-section 500~$\mu$m~$\times$~55~$\mu$m and extends from the top to the bottom surface of the glass substrate ($\sim$1~mm thickness). The MZI consists of two 50:50 directional couplers and has two arms of equal geometrical length; one %of them
\cha{waveguide} crosses \cha{perpendicular to}
%orthogonally
the microchannel, while the other passes externally. (c) Top image of the optical-fluidic interface. (d) Picture of the device with several interferometers and microchannels on chip, together with the fiber arrays for coupling input and output light.}
		\label{fig:schematic}
\end{figure*}

%%%%End of the introduction %%%%%%%%%%%%%%%%%%%%%%%%%%%%%%%%%%%%%%%%%%%%%%%%%%%%%%%%%%%%%%%%%%%%%%%%%%%%%%%%%%%%%%%

The device (Fig.~\ref{fig:schematic}) was fabricated by femtosecond laser micromachining in a fused-silica sample to enable the integration of optical waveguides \cite{gattass2008flm,dellavalle2009mpd} and microfluidic channels \cite{marcinkevicius2001fla,vishnubhatla2009scm} in a three-dimensional architecture \cite{osellame2009ioe}: waveguides with slightly elliptical cross section were fabricated by astigmatic shaping of the writing laser beam \cite{osellame2003fwa} so that a small birefringence is induced to preserve linear polarization; the microchannel was fabricated by irradiating a double pyramidal structure followed by etching in a hydrofluoric acid solution in order to have perfectly vertical walls \cite{marcinkevicius2001fla,vishnubhatla2009scm} (see Supplementary Information).%Appendix).

Photon pairs at $\lambda=785$~nm were generated via spontaneous parametric down-conversion (SPDC) in a nonlinear bismuth borate $\mathrm{BiB_3O_6}$ (BiBO) crystal (see Supplementary Information)%Appendix)
 and collected into polarization maintaining fibres (\citeFig{fig:schematic}(a)) . The photon pairs were coupled into the MZI via fiber arrays. Hong-Ou-Mandel (HOM) interference \cite{hong1987mst,MarshallOE2009} at the first directional coupler generates a two photon NOON state \cite{ra-prl-65-1348} $(\ket{20}+\ket{02})/\sqrt{2}$ which is the state we use to probe the sample. Before interfering in the second directional coupler, the sensing mode acquires a relative phase shift in the microchannel that crosses the sensing arm of the MZI.

Output photons are collected by an array of standard telecommunication fibres (monomodal at 1550~nm wavelength, but multimodal at the 785~nm wavelength used), in order to increase collection efficiency, and detected by four single-photon avalanche photodiodes (A, B, C and D in \citeFig{fig:schematic}(a)), after non-deterministic separation at two 50:50 fiber-splitters. With this detection scheme we are able to monitor the different two photon components of the output state $\ket{11}$, $\ket{20}$ and $\ket{02}$ and renormalize the measured fringes with respect to drifts during the measurement in the coupling between fiber arrays and MZI and source pair production rate (see Supplementary Information).

\begin{figure}
		\centering
		\includegraphics[width=8cm]{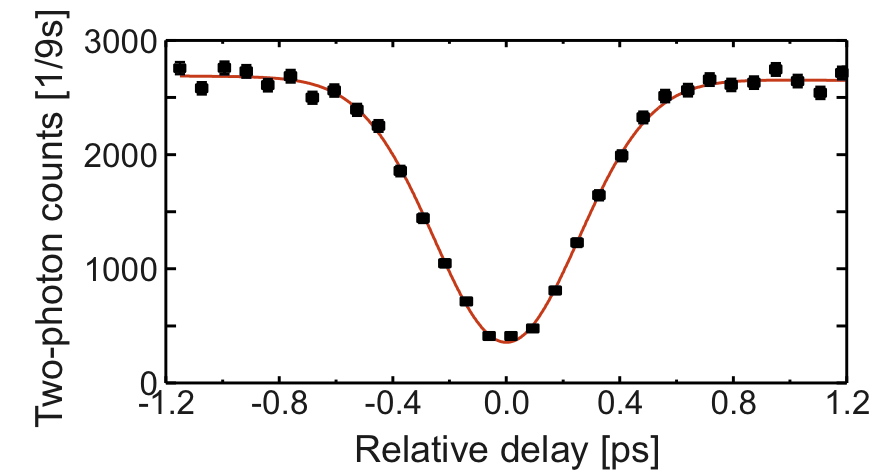}
		\caption{Quantum interference in the Mach-Zehnder interferometer when the microfluidic channel is filled with distilled water. The coincidences at the detectors A and B are plotted as a function of the relative delay between the two photons.}
		\label{fig:hom}
\end{figure}

To establish the quality of quantum interference in our device, we performed a HOM experiment \cite{hong1987mst}, filling the microchannel with distilled water. A lossless MZI composed of two 50:50 couplers, with a relative phase shift $\phi$ between the two arms, is equivalent to a beamsplitter of reflectivity $\mathrm{cos}^2(\phi/2)$. However, asymmetric losses between the interferometer arms limits the maximum visibility to:
\begin{equation}
V_{HOM}=\frac{4T-(T-1)^2}{(T+1)^2}
\label{eq.vis}
\end{equation}
where $T$ is the transmissivity across the microchannel, equivalent to the ratio of the transmissivity of the two arms and $\phi=\pi/2$.

Figure~\ref{fig:hom} shows a typical two-photon detection rate across the two output modes as a function of the relative arrival time, controlled with a translation stage (\citeFig{fig:schematic}(a)). The visibility $V=86.7 \pm 1.3 \%$ is almost ideal for \cha{a ratio of transmissivity} T=61\% since an upper bound of 88\% is calculated from Eq.~\ref{eq.vis}. 
We note that it is merely a coincidence that insertion of distilled water at room temperature in the channel results in $\phi=\pi/2$ to within the precision of this measurement.

%%%%%%%%%%%%%%%%%%%%%%%%%%%%%%%%%%%%%%%%%%%%%%%%%%%%%%

\begin{figure}
		\centering
		\includegraphics[width=8cm]{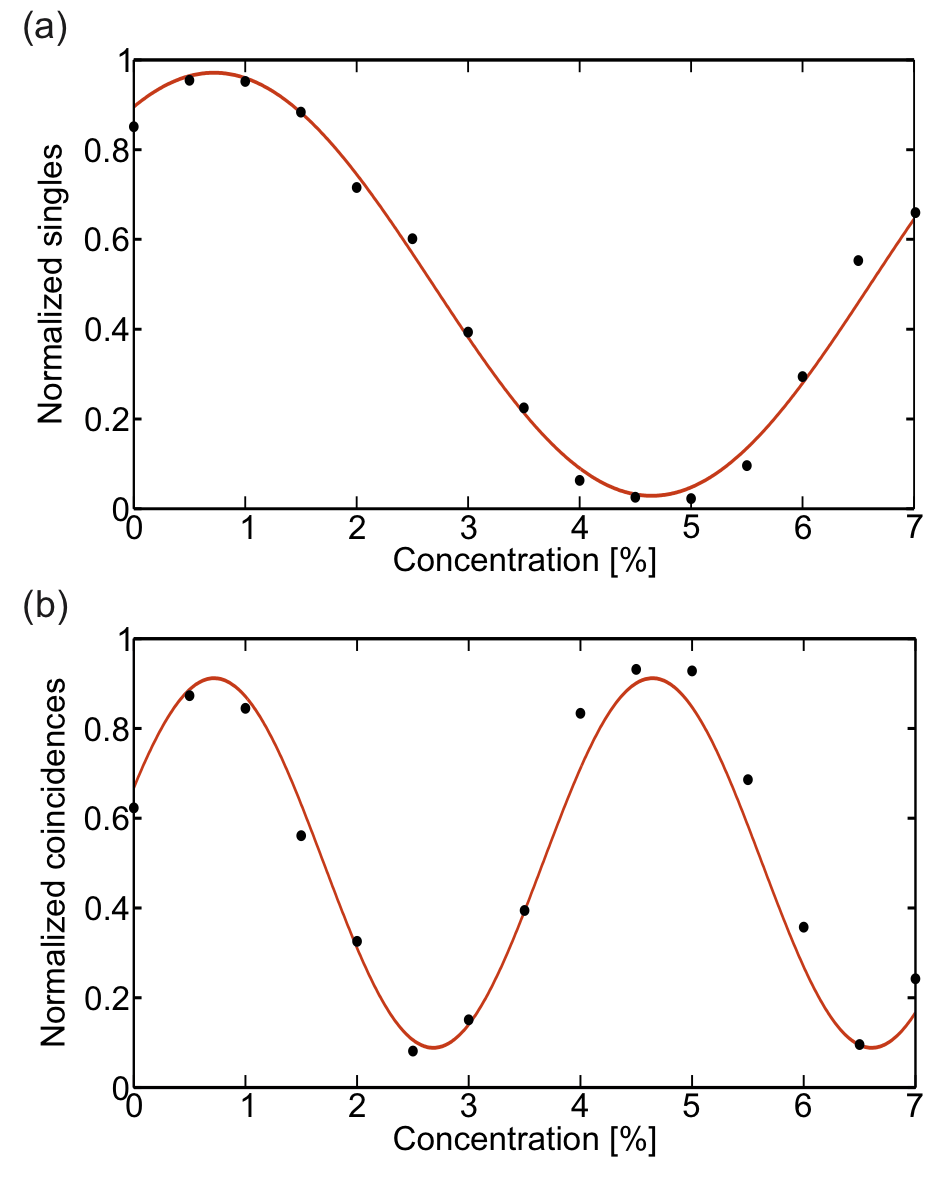}
		\caption{Quantum interference fringes. Normalized single photon counts (a) and two photon coincidences (b) for different concentrations of bovine serum albumin in a buffer solution (full circles). The solid line represent a fitting of the experimental points with a sinusoidal curve. Error bars on data points are the same size of the dots and computed assuming Poissonian statistics of the detection events; other error, arising for example from evaporation, are not taken into account.}
		\label{fig:bsa}
\end{figure}

To test the operation of our device with a real sample we chose Bovine Serum Albumin (BSA) in aqueous buffer solutions as a model fluidic sample that is stable and well-characterized \cite{Peters1997}.
Insertion of the solution in the microchannel was achieved by casting a droplet on the top aperture and exploiting spontaneous filling by capillary action. This geometry is chosen for its simplicity and extension to more sophisticated microfluidic delivery of the solution is straightforward \cite{whitesides2006ofm}.

We performed sensing measurements using one photon and two photon inputs for 15 different concentrations of BSA ranging from 0\% to 7\% in 0.5\% steps. Cleaning of the microchannel with deionized water and acetone was performed before and after each measurement. The single photon fringe is obtained by coupling \cha{only} one photon from the SPDC pair into the MZI and counting the number of detections from one output of the MZI. Figure \ref{fig:bsa}(a) shows the single photon count rate normalized with respect to the sum of the singles from the two outputs (see Supplementary Information) together with theoretical fit function $N_{\ket{10}}=(1+V_{1ph}\mathrm{cos}(\phi+\phi_0))/2$, where $V_{1ph}$ is the fringe visibility, $\phi=\alpha C$ (with $C$ concentration of BSA and $\alpha$ constant), and $\phi_0$ is a constant phase offset term. A visibility $V_{1ph}$=94$\pm 2.2 \%$ is estimated from the fit, compared to a theoretical prediction $V_{theory,1ph}$=97\%, calculated taking into account the device losses.

Two photon fringes were measured by coupling photon pairs into the two input waveguides of the MZI and detecting coincidences from the two separate output channels. Coincidence events C$_{\ket{11}}$ are normalized with respect to the sum of all the possible two photon outputs (see Supplementary Information) %Appendix)
 C$_{\ket{11}}$+C$_{\ket{20}}$+C$_{\ket{02}}$ and shown in \citeFig{fig:bsa}(b) together with the theoretical fit function
\begin{equation}\label{eq.2phfrince}
N_{\ket{11}}=\frac{(1+V_{2ph}\textrm{cos}(2\phi+\tilde{\phi}_0))}{2},
\end{equation}
where the period is half that of the single photon fringe due to super-resolution. The visibility of the fit is $V_{2ph}$=82$\pm 4.8 \%$, in agreement with the theoretical prediction for the interferometer including losses $V_{theory,2ph}$=88\%. This value exceeds the threshold for supersensitivity \cite{re-prl-98-223601,okamoto2008bsq}:
$V_{2ph}^{SQL}=70.7$\%. Current source and detector efficiencies prevent beating the standard quantum limit \cite{walmsley2010arXiv1012.0539D}.
In addition to the 2.1dB ($T=61$\%) loss across the microfluidic channel we estimate 0.5~dB propagation losses and 0.5~dB bending losses, which further increase the visibility threshold.

%%%%%%%%%%%%%%%%%%%%%%%%%%%%%%%%%%%%%%%%%%%%%%%%%%%%%%%%%%%%%%%%%%
\begin{figure}
		\centering
		\includegraphics[width=8cm]{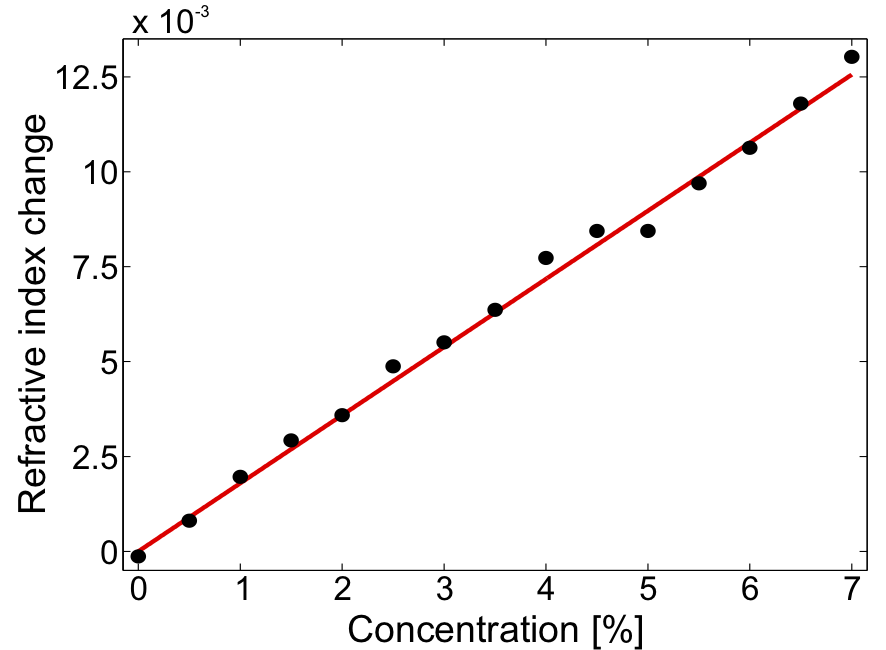}
		\caption{Refractive index change in the buffer solution as a function of BSA concentration. Experimental data (dots) are shown together with a linear fitting (solid curve).}
		\label{fig:Dn}
\end{figure}

The refractive index change $\Delta n_s$ of the BSA solution can be related to the phase shift $\phi$, acquired by light during propagation in the sample, according to:
\begin{equation}\label{eq.dn}
	\Delta n_s= \frac{\lambda}{2 \pi L} \phi
\end{equation}
where $L = 55~\mu m$ is the microchannel length and $\lambda$~=~0.785~$\mu$m is the wavelength.
The dependence of $\Delta n_s$ on the BSA concentration $C$ (\citeFig{fig:Dn}) can thus be inferred from the two-photon coincidences reported in \citeFig{fig:bsa}. The experimental points are well fitted by a linear function, whose slope $dn_s/dC=1.79 \pm 0.04 \times 10^{-3}$ is in very good agreement with the value of $1.82 \times 10^{-3}$, previously reported \cite{barer1954ric} at $\lambda$~=~0.578~$\mu m$.

%%%%%%%%%Conclusions%%%%%%%%%%%%%%%%%%%%

Measurement of the concentration of a protein in solution with entangled states in an integrated quantum photonics device shows the potential for quantum interferometric measurement of light sensitive samples. 
Heralded or deterministic generation of larger entangled states will enable greater sensitivity \cite{ca-prl-99-163604}, when combined with high efficiency photon sources and detectors. Quantum optical circuits of that herald the generation of up to four photon entangled states for quantum metrology have been demonstrated with lithographic waveguides \cite{ma-arxiv-1005.5119}. 
Multipass schemes \cite{hi-nat-450-393} would be compatible with the optofluidic architecture demonstrated here, provided low loss switches can be integrated.
More sophisticated microfluidic delivery systems could be integrated for particular applications \cite{whitesides2006ofm,mo-nphoton-1-106}.
Adding more waveguide capabilities, for example polarization-based quantum measurements \cite{mitchell2004pol, sa-prl-105-200503}, would enable measurement of samples that induce a concentration dependent rotation of the probing light polarization.

%

%\bibliography{QuantumBioSensor_Nat_jm_v2,/Users/job/Dropbox/2011/papers/bib15b.bib}
%\bibliography{QuantumBioSensor_Nat_ml,/Users/job/Dropbox/2011/papers/bib16.bib}
\bibliography{QuantumBioSensor_arxiv}
\section*{APPENDIX}

\noindent\textbf{Fabrication of the optofluidic device:}
The integrated optofluidic device is fabricated using a Yb:KYW cavity-dumped mode-locked oscillator, which delivers 300 fs pulses at 1030 nm wavelength, with 1 MHz repetition rate. The structures are fabricated by direct laser inscription with sample translation enabled by high precision three-axes air bearing stages (Aerotech Fiber-Glide 3D). Laser irradiation for both interferometers and microchannels is performed in the same run. Several devices, each composed of a microchannel interfaced with a Mach-Zehnder interferometer, are realized on the same 2 cm $\times$ 2 cm glass chip.

Waveguides are fabricated using five overlapped scans at 1.5 mm/s with 300 nJ pulses at the second harmonic of the laser (515 nm wavelength), focused at 400 $\mu$m depth into the substrate by a 0.6 NA microscope objective. A cylindrical telescope, shaping the writing beam, is employed to control the shape of the waveguide cross-section \cite{osellame2003fwa}.

\begin{figure}
		\centering
		\includegraphics[width=0.8\columnwidth]{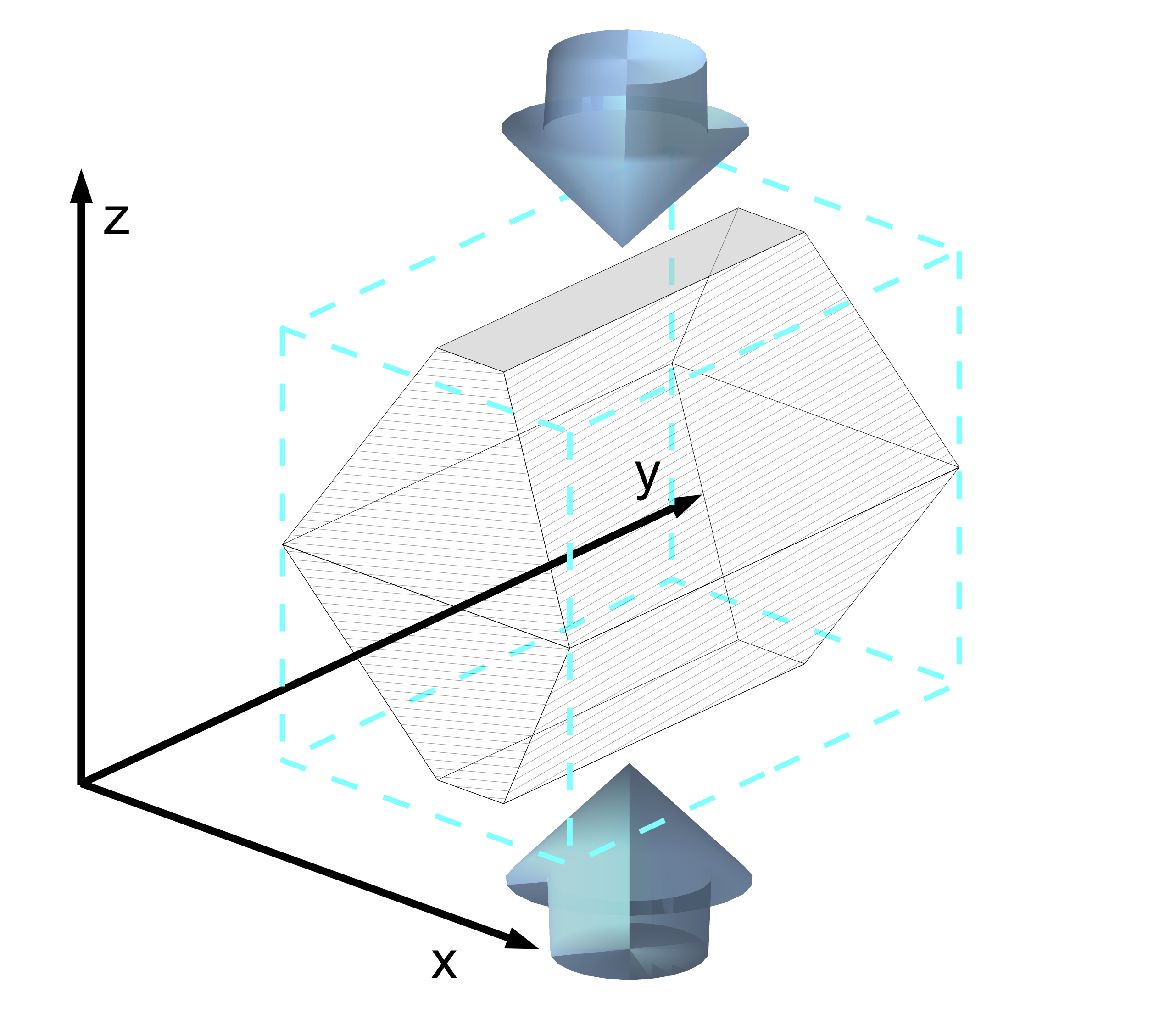}
		\caption{Schematic of the femtosecond laser irradiated path for the fabrication of the microchannel. The double-pyramidal shape is adopted to obtain perfectly vertical walls after the etching process, which proceeds from the top and bottom surfaces of the chip as indicated by the blue arrows. The final shape of the microchannel is depicted by the dashed box.}
		\vspace{-12pt}
		\label{fig:microCHstructure}
\end{figure}

The microchannel is realized using the technique described by \textit{Vishnubhatla et al.}\cite{vishnubhatla2009scm}. In order to obtain perfect verticality of the microchannel walls we need to compensate for the intrinsic tapering induced by the etching process along the z-axis (Fig.\ref{fig:microCHstructure}). This is achieved by irradiating a helicoidal path in order to obtain double-pyramidal structure. A single laser scan at 0.5 mm/s, using 300 nJ pulses of the second harmonic, is employed, with the same focusing set-up used for the waveguides. After laser irradiation the sample is immersed in HF solution for 70 min. The HF acid selectively etches the irradiated material, proceeding along the z-axis from the bottom and top surfaces, simultaneously. This technique produces a microchannel through the 1-mm thick glass sample with a uniform and rectangular cross-section of 55 $\mu$m $\times$ 500 $\mu$m. Polishing of the edges of the chip is finally performed, to allow effective fibre butt-coupling to the interferometers.

\vspace{6pt}\noindent\textbf{Photon pair source:}
The reported experiments used degenerate 785 nm photon pairs, generated using spontaneous parametric downconversion in a 2 mm thick, type-I nonlinear bismuth borate $\textrm{BiB}_3\textrm{O}_6$ (BiBO) crystal phase matched for non-collinear generation (3$^{\circ}$ opening angle). The photon pairs were filtered using a 3nm full-width at half-maximum interference filter (IF) and focused with aspheric collecting lenses into polarisation maintaining fibres (see Fig.~1. A of the main text). The down conversion crystal was pumped by a $\sim10$ mw 392.5~nm beam (focused to a waist of $\omega_0 \approx 40$ $\mu$m), generated via collinear up-conversion---using a 2 mm thick BiBO crystal---of an attenuated 785 nm, 180 fs pulsed Ti:sapphire beam. The 392.5 nm light was filtered using four successive dichroic mirrors.

\vspace{6pt}\noindent\textbf{Normalization procedure:}
Both classical and quantum optical interferometric experiments, such as the ones we performed on BSA solutions, generally consist in measuring a specific output state rate. Considering the 2 $\times$ 2-ports Mach-Zehnder interferometer employed in our experiment (Fig. \ref{fig:schematic}(a)), classical measurements can be performed feeding the device with single photons from one input and counting output states  $\ket{01}$ or $\ket{10}$. Quantum measurements with two-photon NOON states can be performed, on the other hand, feeding the interferometer with two indistinguishable photons and detecting the output states $\ket{11}$, $\ket{20}$ or $\ket{02}$. The detection rate of any of these states depends on the interferometric phase (which is what we want to measure), as well as on photon losses and detector efficiencies. If a single output state is monitored, the measurement result is prone to external perturbations, and variations in the losses or in the efficiencies cannot be distinguished from variations in the interferometric phase. Therefore we choose to adopt robust normalization procedures both for classical and two-photon quantum measurements, which correspond to single photons and coincidences detection, respectively.

To build the single photon interference fringe, the following counts are considered:
\begin{eqnarray*}
A1\vert_{\textrm {meas}} = p_{\ket{10}} \eta_A N_1 \\
A2\vert_{\textrm {meas}} = p_{\ket{01}} \eta_A N_2 \\
B1\vert_{\textrm {meas}} = p_{\ket{01}} \eta_B N_1 \\
B2\vert_{\textrm {meas}} = p_{\ket{10}} \eta_B N_2
\end{eqnarray*}
where $Xk\vert_{\textrm {meas}}$ stays for experimental single count rates at detector $X$ when photons are injected into input $k$; $p_{\ket{ij}}$ is the probability to find a $\ket{ij}$ state at the interferometer output, $N_k$ is the rate at which input photons are effectively injected in the device and $\eta_X$ is the efficiency with which a state $\ket{ij}$ is actually detected, and includes losses, fiber coupler splitting ratio and detector efficiency. The dependency from the unknown parameters, $N_k$ and $\eta_X$, can be cancelled by retrieving $p_{\ket{01}}$ from the experimental count rates according to the following formula:
\begin{equation} \label{normsingles}
p_{\ket{01}}=\frac{\sqrt{A2\vert_{\textrm {meas}} B1\vert_{\textrm {meas}}}}{\sqrt{A2\vert_{\textrm {meas}} B1\vert_{\textrm {meas}}} + \sqrt{A1\vert_{\textrm {meas}} B2\vert_{\textrm {meas}}}}
\end{equation}
The value of $p_{\ket{01}}$ can thus be adopted as the normalized quantity for constructing the interference fringe.

An analogous normalization method is employed for the evaluation of the two-photon interference fringe. Coincidence counts at detectors AB, CD, AC and BD are acquired, which can be written as:
\begin{eqnarray*}
AB\vert_{\textrm {meas}} = N_{\ket{11}} \eta_A \eta_B \\
CD\vert_{\textrm {meas}} = N_{\ket{11}} \eta_C \eta_D \\
AC\vert_{\textrm {meas}} = N_{\ket{20}} \eta_A \eta_C \cdot 2 \\
BD\vert_{\textrm {meas}} = N_{\ket{02}} \eta_B \eta_D \cdot 2
\end{eqnarray*}
where $XY\vert_{\textrm {meas}}$ is the experimental coincidence count rate for the couple of detectors XY and $N_{\ket{ij}}$ is the rate of the $\ket{ij}$ state at the output of the interferometer. The factor 2 in the expressions of $AC\vert_{\textrm {meas}}$ and $BD\vert_{\textrm {meas}}$ takes into account that the two photons can come from two equivalent paths. Also in this case we introduce a normalized quantity $p_{\ket{11}}$, which is independent from losses and efficiencies, defined as:
\begin{equation}
p_{\ket{11}}=\frac{N_{\ket{11}}}{N_{\textrm{tot}}}=\frac{N_{\ket{11}}}{N_{\ket{11}}+N_{\ket{20}}+N_{\ket{02}}}
\end{equation}
where $N_{\ket{20}}=N_{\ket{02}}$, when a MZI with two 50:50 couplers is used. The former expression can be retrieved from the acquired coincidence counts according to:
\begin{equation} \label{normx}
p_{\ket{11}}=\frac{\sqrt{AB\vert_{\textrm {meas}} CD\vert_{\textrm {meas}}}}{\sqrt{AB\vert_{\textrm {meas}} CD\vert_{\textrm {meas}}} + \sqrt{AC\vert_{\textrm {meas}} BD\vert_{\textrm {meas}}}}
\end{equation}

\begin{figure}[t]
		\centering
		\includegraphics[width=0.9\columnwidth]{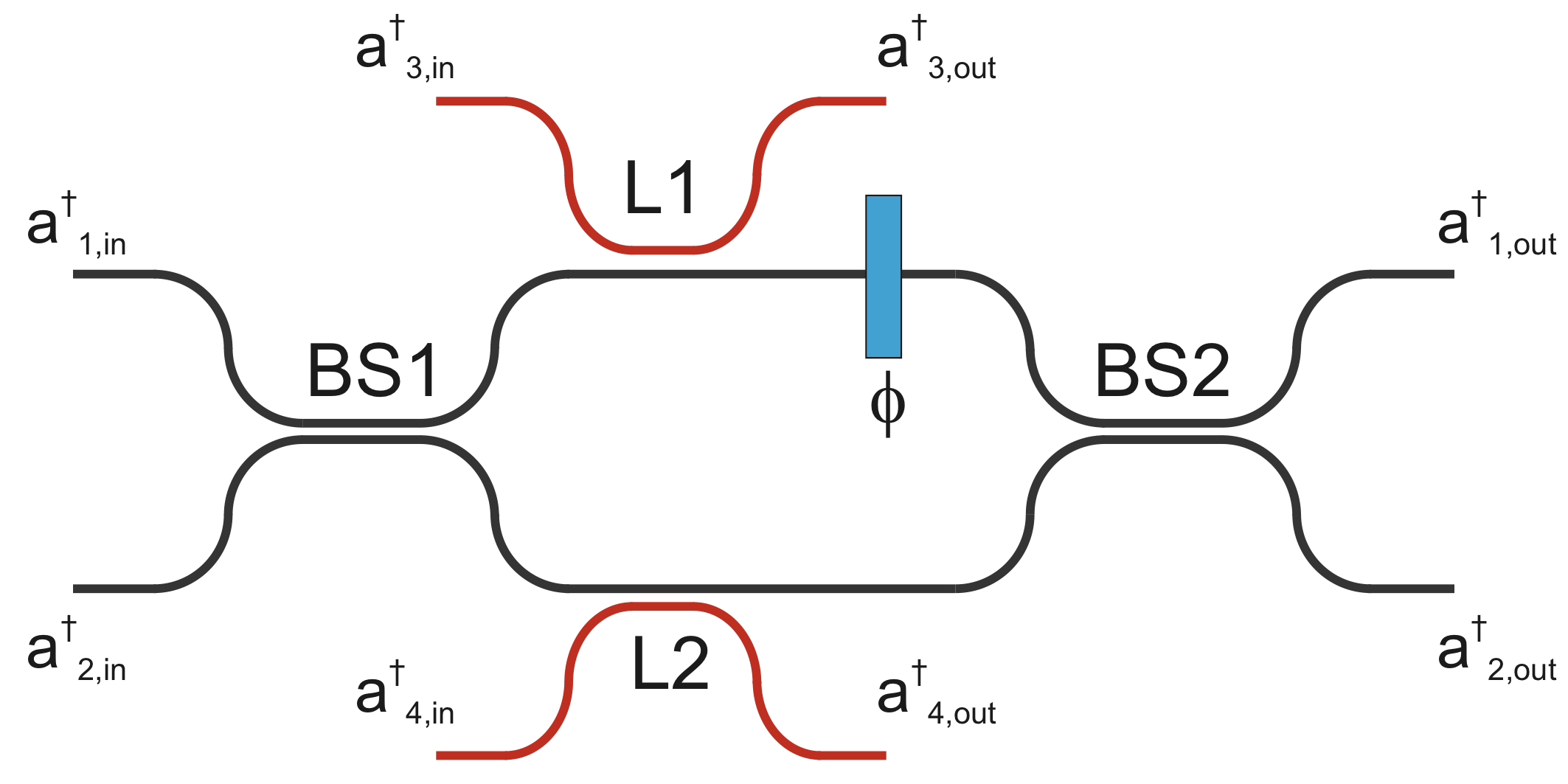}
		\caption{Schematic of the Mach-Zehnder interferometer. Linear losses in the interferometer arms are modeled as extra beamsplitter (in red).}
		\label{fig:SOM1}
\end{figure}

As detailed above, all the experimental quantities $Xk\vert_{\textrm {meas}}$ or $XY\vert_{\textrm {meas}}$ depend on losses and efficiency terms, which in principle could be characterized in each set-up, but in real applications could anyway vary, e.g. due to misalignments induced by external perturbations, thus introducing unpredictable fluctuations in the measurements. On the contrary, the normalized quantities of Eq. \ref{normsingles} and \ref{normx} completely avoid this issue.

\begin{figure}[t!]
		\centering
		\includegraphics[width=0.8\columnwidth]{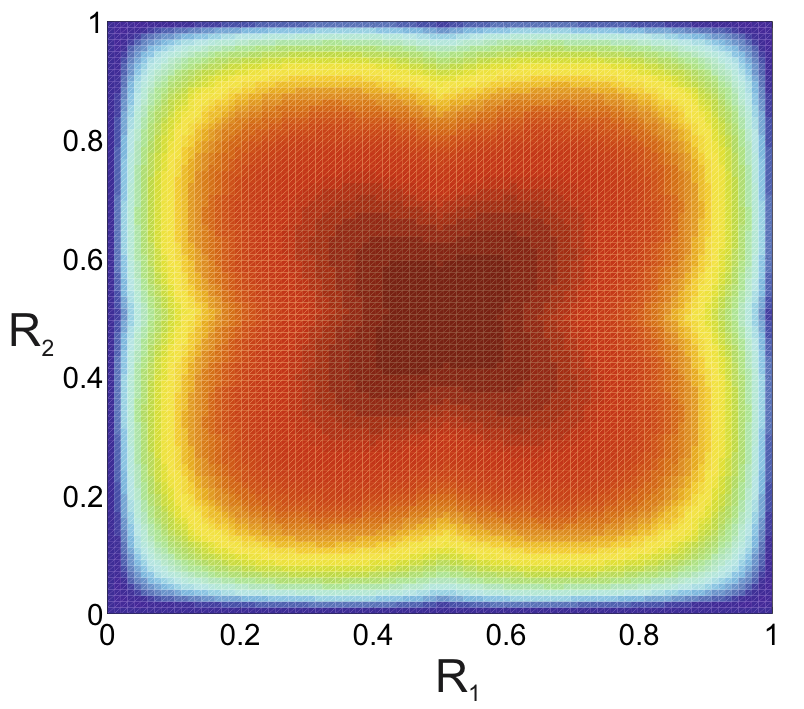}
		\caption{Phase sensitivity $S$ calculated from Eq. \ref{eq.Sens} as a function of the reflection coefficient of the two couplers.}
		\label{fig:SOM2}
\end{figure}

\vspace{6pt}\noindent\textbf{Mach-Zehnder optimal configuration:}
For a  Mach-Zehnder interferometer (MZI) with losses, the probability of a 2-photon coincidence can be expressed as\cite{okamoto2008bsq}
\begin{equation}\label{eq.2phfringe}
P_{11}(\phi)=\eta\frac{(1+V_{2ph}cos(2\phi+\tilde{\phi}_0))}{2}.
\end{equation}
where both the visibility $V_{2ph}$ and the coefficient $\eta$ depend on the losses in the two arms of the interferometer. From this equation we can compute the variance in the coincidences probability as
\begin{equation}\label{eq.2phVar}
\Delta P_{11}(\phi)^2=P_{11}(\phi)(1-P_{11}(\phi))
\end{equation}
while the actual variance of the phase estimation is given by:
\begin{equation}\label{eq.2phfringe}
\delta\phi^2=\frac{\Delta P_{11}(\phi)^2}{(\frac{dP_{11}(\phi)}{d\phi})^2}.
\end{equation}
In order to estimate the probability $P_{11}(\phi)$ we analyse the MZI with losses in the Heisenberg picture. Linear losses in the interferometer arms are modelled as directional couplers where vacuum is injected (Fig.\ref{fig:SOM1}) in one of the inputs. These ancillary modes are labelled $\hat{a}^\dag_3$ and $\hat{a}^\dag_4$ while the real inputs of the MZI are $\hat{a}^\dag_{1,in}$ and $\hat{a}^\dag_{2,in}$.

We will now calculate the probability of a coincidence event as a function of the reflectivity of the two directional couplers of the interferometer, considering the losses in the two arms as parameters. Assuming that the first directional coupler has a reflectivity $r_1$, with $t_1=\sqrt{1-r^2_1}$, it will transform the creation operators of the input modes according to
\begin{equation}\label{eq.BS1}
\dbinom{\hat{a'}^\dag_1}{\hat{a'}^\dag_2}=\dbinom{r_1\hat{a}^\dag_{1,in}+jt_1\hat{a}^\dag_{2,in}}{jt_1\hat{a}^\dag_{1,in}+r_1\hat{a}^\dag_{2,in}}
\end{equation}
The supplementary couplers have transmissivity coefficients that are related to the channel losses by $L_i=t_{L_i}^2$, therefore the linear losses transform the creation operators into
\begin{equation}\label{eq.Losses}
\dbinom{\hat{a''}^\dag_1}{\hat{a''}^\dag_2}=
\dbinom{r_{L_1}\hat{a'}^\dag_{1}+jt_{L_1}\hat{a}^\dag_{3,in}}{r_{L_2}\hat{a'}^\dag_{2}+jt_{L_2}\hat{a}^\dag_{4,in}}.
\end{equation}
The phase shifting operator acting on the sensing arm gives:
\begin{equation}\label{eq.Phase}
\dbinom{\hat{a'''}^\dag_1}{\hat{a'''}^\dag_2}=
\dbinom{\hat{a''}^\dag_{1}e^{i\phi}}{\hat{a''}^\dag_{2}}
\end{equation}
while the second beam splitter generates the final output given by:
\begin{widetext}
\begin{eqnarray}\label{eq.BS2}
&&\dbinom{\hat{a}^\dag_{1,out}}{\hat{a}^\dag_{2,out}}=
\dbinom{r_2\hat{a'''}^\dag_{1}+jt_2\hat{a'''}^\dag_{2}}{jt_2\hat{a'''}^\dag_{1}+r_2\hat{a'''}^\dag_{2}}=
\\
&&\dbinom{(r_{L_1}r_1r_2e^{j\phi}-r_{L_2}t_1t_2)\hat{a}^\dag_{1,in}+j(r_{L_1}t_1r_2e^{j\phi}+r_{L_2}r_1t_2)\hat{a}^\dag_{2,in}
+jt_{L1}r_2e^{j\phi}\hat{a}^\dag_{3,in}-t_{L_2}t_2\hat{a}^\dag_{4,in}}
{j(r_{L_1}r_1t_2e^{j\phi}+r_{L_2}t_1r_2)\hat{a}^\dag_{1,in}-(r_{L_1}t_1t_2e^{j\phi}-r_{L_2}r_1r_2)\hat{a}^\dag_{2,in}
-t_{L1}t_2e^{j\phi}\hat{a}^\dag_{3,in}+jt_{L_2}r_2\hat{a}^\dag_{4,in}}\nonumber
\end{eqnarray}
\end{widetext}
From Eq. \ref{eq.BS2} it is possible to calculate the probability of a coincidence event as a function of the splitting ratio of the MZI directional couplers, which is given by:
\begin{widetext}
\begin{equation}\label{eq.P11}
P_{11}(\phi)=|2r_1t_1r_2t_2(r^2_{L_1}e^{2j\phi}+r^2_{L_2})\\+r_{L_1}r_{L_2}(1-4r^2_1r^2_2+2(r^2_1+r^2_2))e^{j\phi}|^2
\end{equation}
\end{widetext}
Once the coincidence probability is known it is possible to calculate the maximum sensitivity $S$ of the device for a two photon measurement given by the square root of the inverse of the phase variance:
\begin{equation}\label{eq.Sens}
S^2=max_\phi(\delta\phi^2)^{-1}=max_\phi(\frac{(\frac{dP_{11}(\phi)}{d\phi})^2}{\Delta P_{11}(\phi)^2}).
\end{equation}
Figure \ref{fig:SOM2} shows the phase sensitivity $S$ as function of the reflection coefficients $R_i=r_i^2$ of the two directional couplers of the interferometer with a maximum for the case where $R_1=R_2=0.5$.

%\bibliographystyle{unsrt}
%\bibliography{QuantumBioSensor_Nat_v1}

%\begin{thebibliography}{1}

%\bibitem{osellame2003fwa}
%R.~Osellame, S.~Taccheo, M.~Marangoni, R.~Ramponi, P.~Laporta, D.~Polli,
%  S.~De~Silvestri, and G.~Cerullo.
%%\newblock Femtosecond writing of active optical waveguides with astigmatically shaped beams.
%\newblock {\em J. Opt. Soc. Am. B}, 20(7):1559--1567, 2003.

%\bibitem{vishnubhatla2009scm}
%K.C. Vishnubhatla, N.~Bellini, R.~Ramponi, G.~Cerullo, and R.~Osellame.
%%\newblock {Shape control of microchannels fabricated in fused silica by femtosecond laser irradiation and chemical etching}.
%\newblock {\em Optics Express}, 17(10):8685--8695, 2009.
%
%\bibitem{okamoto2008bsq}
%R.~Okamoto, H.F. Hofmann, T.~Nagata, J.L. O'Brien, K.~Sasaki, and S.~Takeuchi.
%%\newblock {Beating the standard quantum limit: phase super-sensitivity of N-photon interferometers}.
%\newblock {\em New Journal of Physics}, 10:073033, 2008.

%\end{thebibliography}

\end{document}
\end{document